\documentclass[10pt]{article}

\usepackage[dvips]{graphicx}
\usepackage{epic}
\usepackage{curves}
\usepackage{amsfonts}

\newcommand{\bela}[1]{\begin{equation}\label{#1}}
\newcommand{\ela}{\end{equation}}
\newcommand{\bear}[1]{\begin{array}{#1}}
\newcommand{\ear}{\end{array}}

\renewcommand{\Psi}{\mbox{\boldmath $\psi$}}

\newcommand{\as}{\\[.6em]}
\newcommand{\As}{\\[.9em]}
\newcommand{\AS}{\\[1.2em]}

\newcommand{\dis}{\displaystyle}

\newcommand{\eins}{\mbox{\rm 1\hspace{-2.2pt}l}}

\newcommand{\tra}{^{\scriptscriptstyle \mathsf{T}}}

\newcommand{\up}[1]{\vec{#1}}
\newcommand{\down}[1]{\underline{#1}}
\newcommand{\both}[1]{\vec{\underline{#1}}}
\newcommand{\vier}[1]{\mbox{\Large$<$}#1\mbox{\Large$>$}}
\newcommand{\drei}[1]{\mbox{\Large$<$}#1|}

\newtheorem{theorem}{Theorem}

\newtheorem{definition}{Definition}

\begin{document}
\begin{center}
  \Large\bf
  On the unification of classical and novel integrable surfaces:\\
  II.\ Difference geometry\\[8mm]
  \large\sc By W.K. Schief\\[2mm]
  \small\sl School of Mathematics, The University of New South Wales,\\
  Sydney, NSW 2052, Australia\\[9mm]
\end{center}

\begin{abstract}
A novel class of discrete integrable surfaces is recorded. This class of 
discrete O~surfaces is shown to include discrete analogues of classical 
surfaces such as isothermic, `linear' Weingarten, Guichard and Petot 
surfaces. Moreover, natural discrete analogues of the 
Gau{\ss}ian and mean curvatures for surfaces parametrized in terms of 
curvature coordinates are used to define surfaces of constant discrete 
Gau{\ss}ian and mean curvatures and discrete minimal surfaces. Remarkably, 
these turn out to be proto-typical examples of discrete O surfaces. 
It is demonstrated that the construction of a 
B\"acklund transformation for discrete O surfaces leads in a natural manner to 
an associated parameter-dependent linear representation. Canonical 
discretizations of the classical pseudosphere and breather pseudospherical
surfaces are generated. Connections with
pioneering work by Bobenko and Pinkall are established.
\end{abstract}

\section{Introduction}

In Schief \& Konopelchenko (2000), a novel class of integrable surfaces has
been introduced and shown to include as canonical members a variety of 
classical surfaces such as isothermic, constant mean curvature, minimal, 
`linear' Weingarten, Guichard and Petot surfaces and surfaces of constant 
Gau{\ss}ian curvature. The definition of this class of O surfaces is
entirely based on the classical notions of conjugate and curvature
coordinates. Both coordinate systems possess natural difference-geometric
counterparts which have been widely used in the construction of integrable 
discrete geometries (Bobenko \& Seiler 1999). It turns out that these may 
indeed be used to define discrete O surfaces. In particular, 
difference-geometric analogues of the above-mentioned classical surfaces
are readily constructed. Thus, the formalism developed in this paper may be
regarded as a first step towards a unified description of 
integrability-preserving discretizations of differential geometries.

Here, we consider two-dimensional
lattices (discrete surfaces) in a Euclidean space
$\mathbb{R}^3$ which 
consist of planar quadrilaterals. These conjugate lattices may
be mapped to parallel conjugate lattices by means of the discrete 
Combescure transformation (Konopel\-chenko \& Schief 1998). Moreover, if
a conjugate lattice is discrete orthogonal, that is the quadrilaterals are
inscribed in circles (Bobenko \& Seiler~1999), 
then the lattice gives rise to a one-parameter family
of parallel lattices on the unit sphere (Konopelchenko \& Schief 1998). As an 
important preliminary, we
exploit the existence of these `spherical representations' to define in a
geometric and algebraic manner discrete Gau{\ss}ian and mean curvatures for
curvature lattices. Sets of~$n$ parallel conjugate lattices 
$\mathbb{R}^3$ are then canonically associated with three parallel conjugate 
lattices in a dual \mbox{(pseudo-)Euclidean} space $\mathbb{R}^n$. If the
dual lattices are also discrete orthogonal then the surfaces in 
$\mathbb{R}^3$ and $\mathbb{R}^n$ are termed (dual) discrete O surfaces.

On appropriate specification of the dimension and the metric of the dual space,
important examples of discrete O surfaces may be isolated. Thus, discrete
isothermic surfaces are obtained which, remarkably, turn out to be precisely 
those proposed by Bobenko \& Pinkall (1996). Moreover, we
show that curvature lattices of constant discrete mean curvature likewise
constitute discrete O surfaces. In fact, discrete constant mean curvature and
mininal surfaces turn out to be particular discrete isothermic surfaces. It is
also demonstrated that any discrete constant mean curvature surface may be
associated with a second parallel discrete constant mean curvature surface
and a parallel 
surface of constant postive discrete Gau{ss}ian curvature, the latter
being another O surface. This result may be interpreted as the analogue
of a classical theorem due to Bonnet and is in agreement with that set down
in Bobenko \& Pinkall (1999). 

Discretizations of Guichard, `linear' Weingarten and Petot surfaces are also
recorded. The discrete Petot surfaces are shown to coincide with
those defined and studied in Schief (1997).
A B\"acklund transformation for discrete O surfaces is obtained by 
constraining the discrete analogue of the classical 
Fundamental transformation~(Konopelchenko \& Schief 1998) 
in such a way that the discrete 
orthogonality conditions are preserved. As a by-product, a matrix Lax pair for
discrete O~surfaces is derived. 
As an application of the B\"acklund 
transformation for discrete O~surfaces, discretizations of the classical 
pseudosphere and breather pseudospherical surfaces are generated.

\section{Conjugate lattices and the discrete Combescure transformation}

In the following, we are concerned with the geometry of two-dimensional
lattices in a three-dimensional Euclidean space, that is maps
\bela{E1}
  \up{R} : \mathbb{Z}^2\rightarrow\mathbb{R}^3,\quad 
  (n_1,n_2)\mapsto \up{r}(n_1,n_2).
\end{equation}
These may be regarded as difference-geometric analogues or {\em 
discretizations} of surfaces
in $\mathbb{R}^3$ and are therefore referred to as {\em discrete surfaces}
(Bobenko \& Seiler~1999). If the position vector~$\up{R}$ to a discrete surface
$\Sigma\subset\mathbb{R}^3$ obeys a `hyperbolic' linear difference equation
of the form
\bela{E2}
  \up{R}_{(12)}-\up{R} = a(\up{R}_{(1)}-\up{R}) + b(\up{R}_{(2)}-\up{R}),
\end{equation}
where the notation 
\bela{E3}
  \bear{rlrl}
    \up{R} = & \up{R}(n_1,n_2),\quad &\up{R}_{(12)} = &\up{R}(n_1+1,n_2+1)\as
    \up{R}_{(1)} = & \up{R}(n_1+1,n_2),\quad & 
    \up{R}_{(2)} = & \up{R}(n_1,n_2+1)
  \ear
\end{equation}
has been adopted, then the lattice is termed {\em conjugate} 
(Bobenko \& Seiler 1999). In geometric terms, this algebraic condition is 
equivalent to the requirement that the quadrilaterals 
$\vier{\up{R},\up{R}_{(1)},\up{R}_{(2)},\up{R}_{(12)}}$ 
of the lattice $\Sigma$ be planar. In this case, it is natural to introduce
the decomposition
\bela{E4}
  \up{R}_{(1)} - \up{R} = \up{X}H,\quad \up{R}_{(2)} - \up{R} = \up{Y}K.
\end{equation}
Since the `tangent vectors' $\up{X}$ and $\up{Y}$ are only defined up to
their moduli, one is at liberty to choose a convenient `gauge'. Indeed,
we here repair to the gauge employed in Konopelchenko \& Schief (1998). Thus,
the compatibility condition for the relations (\ref{E4}) is readily shown to
lead to the linear systems
\bela{E5}
  \bear{rlrl}
    \up{X}_{(2)} = & \dis \frac{\up{X} + q\up{Y}}{\Gamma},\quad &
    H_{(2)} = & \dis \frac{H + pK}{\Gamma}\AS
    \up{Y}_{(1)} = & \dis \frac{\up{Y} + p\up{X}}{\Gamma},\quad &
    K_{(1)} = & \dis \frac{K + qH}{\Gamma},
  \ear
\end{equation}
where $\Gamma$ is defined by
\bela{E6}
  \Gamma^2 = 1 - pq
\end{equation}
and the functions $p$ and $q$ are related to $a$ and $b$ by
\bela{E7}
   a = \frac{H_{(2)}}{\Gamma H},\quad b = \frac{K_{(1)}}{\Gamma K}.
\end{equation}
The system (\ref{E5})$_{2,4}$ may be regarded as {\em adjoint} to the
linear system (\ref{E5})$_{1,3}$.

Conversely, if $\{\up{X},\up{Y},H,K\}$ constitutes a solution of the linear
systems (\ref{E5}) for some functions $p$ and $q$ then the relations
(\ref{E4}) are compatible and $\up{R}$ may be interpreted as the position
vector of a conjugate lattice $\Sigma\subset\mathbb{R}^3$. A second solution
$\{H_*,K_*\}$ of the adjoint system (\ref{E5})$_{2,4}$ gives rise to a 
second conjugate lattice $\Sigma_*$, the position vector of which is defined
by
\bela{E8}
  \up{R}_{*(1)} - \up{R}_* = \up{X}H_*,\quad
  \up{R}_{*(2)} - \up{R}_* = \up{Y}K_*.
\end{equation}
Accordingly, corresponding edges of the quadrilaterals 
$\vier{\up{R},\up{R}_{(1)},\up{R}_{(2)},\up{R}_{(12)}}$ and
$\vier{\up{R}_*,\up{R}_{*(1)},
\up{R}_{*(2)},\up{R}_{*(12)}}$ are parallel and hence the quadrilaterals
themselves are parallel. The lattice $\Sigma_*$ is termed a
{\em discrete Combescure transform} (Konopelchenko \& Schief 1998) of the
lattice $\Sigma$. Thus, the discrete Combescure transformation maps
conjugate lattices to {\em parallel} conjugate lattices. 

\section{Curvature lattices and discrete Gau{\ss}ian and mean curvatures}

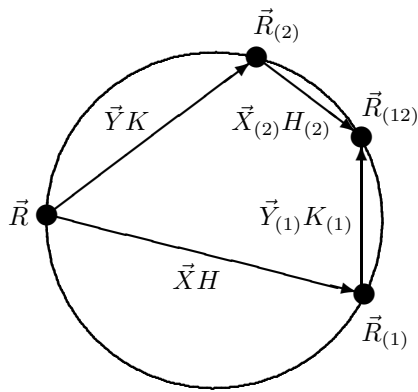
\begin{figure}
\begin{center}
\setlength{\unitlength}{0.00073300in}%
\begin{picture}(2800,2530)(2476,-2897)
\thicklines
\put(2701,-1861){\circle*{150}}
\put(4201,-736){\circle*{150}}
\put(4951,-1306){\circle*{150}}
\put(4971,-2431){\circle*{150}}
\put(2701,-1861){\vector( 4,-1){2220}}
\put(4951,-2386){\vector( 0, 1){1030}}
\put(2701,-1861){\vector( 4, 3){1460}}
\put(3901,-1901){\bigcircle{2404}}
\put(4201,-736){\vector( 4,-3){710}}
\put(2426,-1936){$\up{R}$}
\put(4951,-2761){$\up{R}_{(1)}$}
\put(4951,-1156){$\up{R}_{(12)}$}
\put(4176,-566){$\up{R}_{(2)}$}
\put(3601,-2391){$\up{X}H$}
\put(3111,-1256){$\up{Y}K$}
\put(4021,-1256){$\up{X}_{(2)}H_{(2)}$}
\put(4220,-1936){$\up{Y}_{(1)}K_{(1)}$}
\end{picture}
\end{center}
\caption{An embedded quadrilateral of a curvature lattice}
\label{cyclic}
\end{figure}

Conjugate lattices are said to be {\em curvature lattices} if they are, in
addition, {\em discrete orthogonal}\footnote{It is emphasized that discrete
orthogonality is only defined in conjunction with conjugacy.} 
(Bobenko \& Seiler 1999), that is if the quadrilaterals are inscribed in
circles. Analytically, curvature lattices are characterized by the 
following three equivalent conditions:
\bela{E9}
  \bear{rl}
    \mbox{\sf(i)} \quad & \up{X}_{(2)}\cdot\up{Y} + \up{Y}_{(1)}\cdot\up{X} 
                           = 0\as
   \mbox{\sf(ii)} \quad & 2\up{X}\cdot\up{Y} + p\up{X}^2 + q\up{Y}^2 = 0\as
  \mbox{\sf(iii)} \quad & \up{X}_{(2)}^2 = \up{X}^2,
                       \quad \up{Y}_{(1)}^2 = \up{Y}^2.
  \ear
\end{equation}
In fact, since the tangent vectors $\up{X}$ and $\up{Y}$ of conjugate 
lattices are oriented in such a way that
\bela{E10}
  \up{X}_{(2)}\times\up{Y} + \up{Y}_{(1)}\times\up{X} = 0,
\end{equation}
combination of the conditions {\sf (i)} and {\sf (iii)} shows that opposite
angles in any quadrilateral are either equal or add up to $\pi$ 
according to whether the edges intersect or not. This proves that
the quadrilaterals are inscribed in circles. An `embedded' 
quadrilateral is displayed in Figure \ref{cyclic}.
It is also
noted that the condition {\sf (iii)} implies that we may assume without loss
of generality that $\up{X}$ and $\up{Y}$ constitute unit vectors.
Indeed, it is readily verified that appropriate scaling of 
$\up{X},\up{Y},H,K$ and~$p,q$ leads to 
\bela{E11}
  \up{X}^2=1,\quad \up{Y}^2=1.
\end{equation}
This normalization will be adopted throughout the paper.

It is evident that the Combescure transformation maps within the class of
curvature lattices since the discrete orthogonality conditions (\ref{E9})
do not involve the solutions $\{H,K\}$ of the adjoint system 
(\ref{E5})$_{2,4}$. This implies that any curvature lattice may be mapped
via a Combescure transformation onto the unit sphere $\mathbb{S}^2$. 
In fact, as pointed out in Konopelchenko \& Schief (1998), there exists a
one-parameter family of curvature lattices $\Sigma_{\circ}$ which are parallel
to any given curvature lattice. These are constructed by choosing an arbitrary 
point $\up{R}_{\circ}(0,0)\in\mathbb{S}^2$ corresponding to the vertex
$\up{R}(0,0)$ of the curvature lattice $\Sigma$ and successively drawing lines 
parallel to the edges of~$\Sigma$, thereby identifying the points of 
intersection with 
the unit sphere as the vertices of the parallel lattice $\Sigma_{\circ}$.
It is natural to refer to this family of parallel curvature lattices as 
{\em spherical representations} of the curvature lattice $\Sigma$. Each
spherical representation~$\Sigma_{\circ}$ corresponds to a particular
solution $\{H_{\circ},K_{\circ}\}$ of the adjoint system 
\bela{E12}
  H_{\circ(2)} = \frac{H_{\circ} + pK_{\circ}}\Gamma,\quad
  K_{\circ(1)} = \frac{K_{\circ} + qH_{\circ}}\Gamma.
\end{equation}

We are now in a position to define discrete Gau{\ss}ian and mean
curvatures for conjugate lattices. Thus, let
$\vier{\up{R},\up{R}_{(1)},\up{R}_{(2)},\up{R}_{(12)}}$ be an embedded
quadrilateral of a curvature
lattice $\Sigma$. The area $A_u$ of the `upper' triangle
$\drei{\up{R},\up{R}_{(2)},\up{R}_{(12)}}$ is then readily shown to be
\bela{E13}
  A_u = |H_{(2)}K|\Xi,\quad \Xi = \frac{\sqrt{1
         -(\up{X}\cdot\up{Y})^2}}{2\Gamma}
\end{equation}
while the area $A_l$ of the `lower' triangle 
$\drei{\up{R},\up{R}_{(1)},\up{R}_{(12)}}$ reads
\bela{E14}
  A_l = |K_{(1)}H|\Xi.
\end{equation}
The total area of the quadrilateral is given by
\bela{E15}
  A = |H_{(2)}K + K_{(1)}H|\Xi.
\end{equation}
In a similar manner, one obtains for the total area of the parallel
quadrilateral $\vier{\up{R}_{\circ},\up{R}_{\circ(1)},\up{R}_{\circ(2)},
\up{R}_{\circ(12)}}$ of a spherical representation $\Sigma_{\circ}$ the
expression
\bela{E16}
 A_{\circ} = |H_{\circ(2)}K_{\circ} + K_{\circ(1)}H_{\circ}|\Xi.
\end{equation}
In analogy to the differential-geometric case (Eisenhart 1960) in which the 
Gau{\ss}ian
curvature of a surface $\Sigma$ is defined as the inverse ratio of the 
areas of an infinitesimal
surface element and its spherical representation, that is
\bela{E16a}
  \mathsf{K} = \frac{(\up{N}_x\times\up{N}_y)\cdot\up{N}}{
                     (\up{R}_x\times\up{R}_y)\cdot\up{N}}
             = \pm\frac{|\up{N}_x\times\up{N}_y|}{
                        |\up{R}_x\times\up{R}_y|},
\end{equation}
where $x,y$ and $\up{N}$ denote curvature coordinates and the unit normal to 
$\Sigma$ respectively, we now propose the following:\footnote{The sign in
(\ref{E17}) is chosen in such a way that the classical expression for the
Gau{\ss}ian curvature is obtained in the natural continuum limit.}

\begin{definition} {\bf (Discrete Gau{\ss}ian curvature)} The {\em discrete
Gau{\ss}ian curvature} of a curvature lattice $\Sigma$ with respect to an
associated spherical representation~$\Sigma_{\circ}$ is defined by
\bela{E17}
  \mathsf{K} = \frac{H_{\circ(2)}K_{\circ} + K_{\circ(1)}H_{\circ}}{
                     H_{(2)}K + K_{(1)}H}.
\end{equation}
\end{definition}

\noindent
It is noted that, by construction, the above definition of the discrete
Gau{\ss}ian curvature is `geometric' in the sense that it is invariant under a 
re-labelling of the curvature lattice. Indeed, it is readily verified that 
$\mathsf{K}$ may also be brought into the form
\bela{E18}
  \mathsf{K} = \frac{H_{\circ(2)}K_{\circ(1)}+ H_{\circ} K_{\circ}}{
                     H_{(2)}K_{(1)} + HK}.
\end{equation}

The mean curvature of a surface $\Sigma$ parametrized in terms of curvature
coordinates is given by
\bela{E19}
  \mathsf{M} = -\frac{(\up{N}_x\times\up{R}_y)\cdot\up{N} +
                      (\up{R}_x\times\up{N}_y)\cdot\up{N}}{
                      (\up{R}_x\times\up{R}_y)\cdot\up{N}} = 
               -\frac{\pm |\up{N}_x\times\up{R}_y| 
                      \pm |\up{R}_x\times\up{N}_y|}{
                      |\up{R}_x\times\up{R}_y|},
\end{equation}
where the signs in the second expression for $\mathsf{M}$ have to be chosen
appropriately. Since the expression $|\up{N}_x\times\up{R}_y|/2$ is 
nothing but the area of the triangle spanned by the vectors $\up{N}_x$ and
$\up{R}_y$, its canonical difference-geometric counterpart could be any of the
areas
\bela{E20}
   |H_{\circ(2)}K|\Xi,\quad |K_{(1)}H_{\circ}|\Xi,\quad
   |H_{\circ(2)}K_{(1)}|\Xi,\quad |H_{\circ}K|\Xi.
\end{equation}
Analogously, the area $|\up{R}_x\times\up{N}_y|/2$ of the triangle spannned
by the vectors $\up{R}_x$ and~$\up{N}_y$ gives rise to the discretizations
\bela{E21}
   |H_{(2)}K_{\circ}|\Xi,\quad |K_{\circ(1)}H|\Xi,\quad 
   |H_{(2)}K_{\circ(1)}|\Xi,\quad |HK_{\circ}|\Xi.
\end{equation}
However, any potential candidate for the discrete mean curvature expressed in
terms of the above areas must prove `geometric'. It turns out there
indeed exists a `symmetric' expression which satisfies this requirement.
Thus, the following definition is invariant under a re-labelling of the
curvature lattice:

\begin{definition} {\bf (Discrete mean curvature)} The {\em discrete
mean curvature} of a curvature lattice $\Sigma$ with respect to an
associated spherical representation $\Sigma_{\circ}$ is defined~by
\bela{E22}
  \mathsf{M} = -\frac{H_{\circ(2)}K + K_{(1)}H_{\circ} 
                   + H_{(2)}K_{\circ} + K_{\circ(1)}H}{
                     H_{(2)}K + K_{(1)}H}.
\end{equation}
\end{definition}

We conclude this section with a natural definition of discrete analogues of 
classical surfaces of constant Gau{\ss}ian and mean curvature and minimal 
surfaces:

\begin{definition} {\bf (Surfaces of constant discrete Gau{\ss}ian and mean
curvature and discrete minimal surfaces)}
A curvature lattice $\Sigma$ constitutes $\mathsf{(i)}$ a {\em surface of 
constant discrete Gau{\ss}ian curvature}, $\mathsf{(ii)}$ a {\em surface of 
constant discrete 
mean curvature}, $\mathsf{(iii)}$ a {\em discrete minimal surface} if there
exists an associated spherical representation~$\Sigma_{\circ}$ such that
\bela{E23}
  \mathsf{(i)\quad K}=\mbox{\rm const},\quad
  \mathsf{(ii)\quad M}=\mbox{\rm const},\quad
  \mathsf{(iii)\quad M}=0
\end{equation}
respectively.
\end{definition}
  
\noindent
It is remarkable that these discrete surfaces prove integrable since they
constitute canonical members of the class of discrete O surfaces to be
introduced in the sequel. 
In fact, in the case of surfaces of constant positive discrete Gau{\ss}ian
and constant discrete mean curvature and discrete minimal surfaces, they
coincide with those proposed by Bobenko \& Pinkall (1996, 1999). 

\section{Parallel conjugate lattices and their duals}

As in the differential-geometric case (Schief \& Konopelchenko 2000), 
we now investigate the geometric and 
algebraic properties of sets $\{\Sigma_1,\ldots,\Sigma_n\}$ of discrete
surfaces which are related by discrete Combescure transformations. To this
end, we consider the linear systems
\bela{E24}
  \bear{rlrl}
    \up{X}_{(2)} = &\dis \frac{\up{X}+q\up{Y}}{\Gamma},\quad &
    \down{H}_{(2)} = &\dis \frac{\down{H}+p\down{K}}{\Gamma}\AS
    \up{Y}_{(1)} = &\dis \frac{\up{Y}+p\up{X}}{\Gamma},\quad &
    \down{K}_{(1)} = &\dis \frac{\down{K}+q\down{H}}{\Gamma},
  \ear
\end{equation}
where $\up{X},\up{Y}\in\mathbb{R}^3$ and $\down{H},\down{K}\in\mathbb{R}^n$ are
interpreted as column and row vectors respectively, and define a matrix
$\both{R}\in\mathbb{R}^{3,n}$ via the compatible equations
\bela{E25}
  \both{R}_{(1)} - \both{R} = \up{X}\down{H},\quad
  \both{R}_{(2)} - \both{R} = \up{Y}\down{K}.
\end{equation}
Here, the function $\Gamma$ is defined, as usual, by $\Gamma^2=1-pq$. Thus, the
geometric interpretation given below is immediate:

\medskip
\noindent
{\em The vectors
 $$ \up{R}_{\kappa}\in\mathbb{R}^3,\quad\kappa=1,\ldots,n $$
parametrize parallel conjugate lattices $\Sigma_{\kappa}\subset
\mathbb{R}^3$ with tangent vectors $\up{X}$ and $\up{Y}$.}

\medskip
\noindent
However, since there exists complete symmetry between $\{\up{X},\up{Y}\}$
and $\{\down{H},\down{K}\}$ and the definition of conjugate lattices is in fact
independent of the dimension of the ambient space, the following point of view
is also valid:

\medskip
\noindent
{\em The vectors
 $$ \down{R}^{k}\in\mathbb{R}^n,\quad k=1,2,3 $$
parametrize parallel conjugate lattices $\Sigma^k\subset
\mathbb{R}^n$ with tangent vectors $\down{H}$ and $\down{K}$.}

\medskip
We refer to the discrete surfaces $\Sigma^k$ as {\em dual} to the 
discrete surfaces $\Sigma_{\kappa}$. 
The concept of dual conjugate lattices has been exploited in the context 
of integrable difference geometries by several authors 
(Konopelchenko \& Schief 1998; Doliwa \& Santini 1999). 
As mentioned earlier, we here regard the ambient space 
$\mathbb{R}^3$ as a Euclidean space even though
the generalization to pseudo-Euclidean spaces $\mathbb{R}^3$ and
their higher-dimen\-sional analogues is straightforward. 
By contrast, it turns out pivotal to deal with pseudo-Euclidean
dual spaces~$\mathbb{R}^n$. Thus, we endow
$\mathbb{R}^n$ with the inner product
\bela{E26}
  \down{H}\cdot\down{K} = \down{H}\,\down{K}\tra = \sum_{\kappa,\mu=1}^n
  H_{\kappa}S^{\kappa\mu}K_{\mu},
\end{equation}
where $S=(S^{\kappa\mu})$ is a constant symmetric matrix.

\section{A novel class of discrete integrable surfaces}

\subsection{The geometry of discrete O surfaces}

Since the discrete Combescure transformation maps within the class of
curvature lattices, it is natural to focus on parallel conjugate
lattices $\Sigma^k$ which are dual to a set of parallel curvature lattices
$\Sigma_{\kappa}$. In the generic case, these are not necessarily discrete 
orthogonal. In fact, the discrete orthogonality condition
\bela{E27}
  \down{H}_{(2)}\cdot\down{K} + \down{K}_{(1)}\cdot\down{H} = 0
\end{equation}
or, equivalently,
\bela{E28}
  \down{H}_{(2)}^2 = \down{H}^2,\quad \down{K}_{(1)}^2 = \down{K}^2
\end{equation}
imposes severe constraints on the discrete surfaces $\Sigma_{\kappa}$. Thus,
imposition of the discrete orthogonality condition on the dual lattices leads
to a natural discretization of the integrable class of {\em O surfaces} 
recorded in Schief \& Konopelchenko (2000).

\begin{definition} {\bf (Discrete O surfaces)} Parallel conjugate lattices
$\Sigma_{\kappa}\subset\mathbb{R}^3$ and their duals 
$\Sigma^k\subset\mathbb{R}^n$ are termed {\em (dual) discrete O surfaces} if
both the lattices $\Sigma_{\kappa}$ and~$\Sigma^k$ are discrete orthogonal.
\end{definition}
 
Before we establish the integrability of discrete O surfaces by deriving a
para\-meter-dependent linear representation and an associated B\"acklund
transformation, we demonstrate below how discrete versions of classical
surfaces such as isothermic, constant mean curvature, minimal, `linear'
Weingarten, Guichard and Petot surfaces and surfaces of constant Gau{\ss}ian
curvature may be retrieved as canonical examples of discrete O surfaces. The
significance of the discrete Gau{\ss}ian and mean curvatures defined in
\S 3 is thereby brought to light.

\subsection{Examples}

\subsubsection{Surfaces of constant discrete Gau{\ss}ian curvature} 

As in Schief \& Konopelchenko (2000), we begin with the simplest choice
\bela{E29}
  S = \left(\bear{cc}1&0\\ 0&1\ear\right)
\end{equation}
corresponding to a two-dimensional Euclidean dual space $\mathbb{R}^2$. In this
case, the discrete orthogonality condition (\ref{E27}) takes the form
\bela{E30}
  H_{1(2)}K_1 + K_{1(1)}H_1 + H_{2(2)}K_2 + K_{2(1)}H_2 = 0.
\end{equation}
By virtue of (\ref{E17}), this is equivalent to the requirement that the 
discrete Gau{\ss}ian curvatures of $\Sigma_1$ and $\Sigma_2$ with respect to
one and hence any spherical representation~$\Sigma_{\circ}$ be related by
\bela{E31}
  \mathsf{K}_1 = -\mathsf{K}_2.
\end{equation}
Alternatively, if we consider a pseudo-Euclidean dual space $\mathbb{R}^2$ with
\bela{E32}
  S = \left(\bear{cc}1&0\\ 0&-1\ear\right)
\end{equation}
then the discrete orthogonality condition reads
\bela{E33}
  H_{1(2)}K_1 + K_{1(1)}H_1 = H_{2(2)}K_2 + K_{2(1)}H_2 
\end{equation}
so that
\bela{E34}
  \mathsf{K}_1 = \mathsf{K}_2.
\end{equation}
We therefore conclude that a pair of parallel curvature lattices constitute
discrete O~surfaces if the discrete Gau{\ss}ian curvatures of corresponding
parallel quadrilaterals are of the same magnitude. In particular, if we 
confine the lattice $\Sigma_2$ to the sphere with $\mathsf{K}_2=1$ then the
discrete surface $\Sigma_1$ is of constant discrete Gau{\ss}ian curvature with 
respect
to the spherical representation $\Sigma_{\circ}=\Sigma_2$. Since in the
differential-geometric setting (Schief \& Konopelchenko 2000) the above
analysis has been shown to lead to classical (pseudo)spherical
surfaces, it
has been established that, remarkably, 
the natural discrete analogues of surfaces of
constant Gau{\ss}ian curvature coincide with the surfaces of discrete
constant Gau{\ss}ian curvature defined in \S 3.

\subsubsection{Discrete isothermic and minimal surfaces} 

The choice
\bela{E35}
  S = \left(\bear{cc}0&1\\ 1&0\ear\right)
\end{equation}
leads to the discrete orthogonality condition
\bela{E36}
  H_{1(2)}K_2 + K_{1(1)}H_2 + H_{2(2)}K_1 + K_{2(1)}H_1 = 0
\end{equation}
and the alternative characterization (\ref{E28}) yields
\bela{E37}
  H_1H_2 = \alpha(n_1),\quad K_1K_2 = -\beta(n_2).
\end{equation}
The latter may be used to eliminate the quantities $H_2$ and $K_2$ in
(\ref{E36}) to obtain
\bela{E38}
  (\beta H_{1(2)}H_1 - \alpha K_{1(1)}K_1)(K_{1(1)}H_{1(2)} + H_1K_1)=0.
\end{equation}
If we now assume that the quadrilaterals are embedded and non-degenerate then
the second factor in (\ref{E38}) is non-vanishing and whence
\bela{E39}
    \frac{H_{1(2)}H_1}{K_{1(1)}K_1} = \frac{\alpha}{\beta}
  = \frac{H_{2(2)}H_2}{K_{2(1)}K_2}
\end{equation}
with $\alpha\beta>0$. In terms of the edges of the quadrilaterals, these 
relations translate into
\bela{E40}
  \frac{|\up{R}_{1(12)}-\up{R}_{1(2)}||\up{R}_{1(1)}-\up{R}_1|}{
        |\up{R}_{1(12)}-\up{R}_{1(1)}||\up{R}_{1(2)}-\up{R}_1|}
 =\frac{\alpha}{\beta} 
 =\frac{|\up{R}_{2(12)}-\up{R}_{2(2)}||\up{R}_{2(1)}-\up{R}_2|}{
        |\up{R}_{2(12)}-\up{R}_{2(1)}||\up{R}_{2(2)}-\up{R}_2|}.
\end{equation}
Curvature lattices with {\em cross-ratios} of the form (\ref{E40}) have been
termed {\em discrete isothermic} by Bobenko \& Pinkall (1996).
Thus, the discrete surfaces $\Sigma_1$ and $\Sigma_2$
constitute {\em discrete isothermic surfaces} which are related by the
{\em discrete Christoffel transformation} (Bobenko \& Pinkall 1996)
\bela{E41}
  \up{R}_{2(1)}-\up{R}_2 = \alpha\frac{\up{R}_{1(1)}-\up{R}_1}{
   (\up{R}_{1(1)}-\up{R}_1)^2},\quad
  \up{R}_{2(2)}-\up{R}_2 = -\beta\frac{\up{R}_{1(2)}-\up{R}_1}{
   (\up{R}_{1(2)}-\up{R}_1)^2}.\end{equation}

If we identify $\Sigma_2$ with a spherical representation of $\Sigma_1$ then
$H_2=H_{\circ}$ and $K_2=K_{\circ}$ which, in turn, implies that the
discrete surface $\Sigma_1$ is discrete minimal since
\bela{E42}
  \mathsf{M}_1 = -\frac{H_{2(2)}K_1 + K_{1(1)}H_2 
                   + H_{1(2)}K_2 + K_{2(1)}H_1}{
                     H_{1(2)}K_1 + K_{1(1)}H_1} = 0
\end{equation}
by virtue of (\ref{E36}). Thus, the Christoffel transform of a discrete
minimal surface constitutes a discrete sphere. This fact has been established 
by Bobenko \& Pinkall~(1996) for {\em discrete minimal
isothermic surfaces}. However, it is evident that any discrete minimal
surface as defined in \S 3 is discrete isothermic.

\subsubsection{Surfaces of constant discrete mean curvature and a discrete 
Bonnet theorem} 

In Schief \& Konopelchenko (2000), it has been shown that
classical surfaces of constant mean curvature constitute O surfaces. It turns
out that the {\em discrete isothermic constant mean curvature surfaces}
recorded in Bobenko \& Pinkall (1996) are indeed discrete O surfaces and, in 
fact, coincide with the class of constant discrete mean curvature surfaces
defined in \S 3. Thus, any surface of constant discrete mean curvature
is in fact discrete isothermic as in the differential-geometric context.
In order to establish this result, we first observe that any set of parallel
discrete O surfaces $\Sigma_{\kappa}$ gives rise to an infinite number of
parallel discrete O surfaces by taking linear combinations of the associated
position vectors $\up{R}_{\kappa}$. For instance, if $\Sigma_1$ and $\Sigma_2$
are two discrete isothermic surfaces related by the discrete Christoffel
transformation then the discrete surfaces $\Sigma_{\pm}$ with position vectors
\bela{E43}
  \up{R}_{\pm} = \frac{1}{2}(\up{R}_2\pm\up{R}_1)
\end{equation}
constitute discrete O surfaces which are parallel to both $\Sigma_1$ and  
$\Sigma_2$. The corresponding solutions of the adjoint system 
(\ref{E5})$_{2,4}$ are given by
\bela{E44}
   H_{\pm} = \frac{1}{2}(H_2\pm H_1),\quad K_{\pm} = \frac{1}{2}(K_2\pm K_1).
\end{equation}
Accordingly, the discrete Gau{\ss}ian curvatures of the discrete surfaces 
$\Sigma_{\pm}$ take the form
\bela{E45}
  \mathsf{K}_{\pm} = \frac{H_{\circ(2)}K_{\circ}+K_{\circ(1)}H_{\circ}}{
                           H_{\pm(2)}K_{\pm}+K_{\pm(1)}H_{\pm}} = 
     \frac{4(H_{\circ(2)}K_{\circ}+K_{\circ(1)}H_{\circ})}{
     H_{1(2)}K_1+K_{1(1)}H_1 + H_{2(2)}K_2+K_{2(1)}H_2}
\end{equation}
by virtue of (\ref{E36}) and hence coincide. 
This is not surprising since the transition from  
$(\Sigma_1,\Sigma_2)$ to $(\Sigma_+,\Sigma_-)$ may be interpreted at the
level of the matrix $S$ as a similarity transformation  mapping the case
(\ref{E35}) to the case (\ref{E32}).

If we now identify the discrete surface $\Sigma_-$ with a spherical 
representation $\Sigma_{\circ}$ of the discrete isothermic surfaces, that is
\bela{E46}
   \up{R}_- = \up{R}_{\circ},\quad H_- = H_{\circ},\quad K_- = K_{\circ},
\end{equation}
then
\bela{E47}
  \mathsf{K}_{\pm}=1,\quad \mathsf{M}_1=1,\quad \mathsf{M}_2=-1.
\end{equation}
The latter relations encapsulate a discrete version of a classical theorem 
due to Bonnet (Eisenhart 1960):

\medskip\noindent
{\em With any surface $\Sigma_+$ of 
constant discrete Gau{\ss}ian curvature $\mathsf{K}_+=1$ one may associate two
parallel surfaces $\Sigma_1$ and $\Sigma_2$ of constant discrete mean 
curvature \mbox{$\mathsf{M}_1=1$}
and $\mathsf{M}_2=-1$ respectively with position vectors}
\bela{E48}
  \up{R}_1 = \up{R}_+ - \up{R}_{\circ},\quad 
  \up{R}_2 = \up{R}_+ + \up{R}_{\circ}.
\end{equation}

\noindent   
This is in agreement with a result presented in Bobenko \& Pinkall (1999) and
may be regarded as 
a special case of the following statement:

\medskip\noindent
{\em If the discrete Gau{\ss}ian curvatures of two parallel curvature lattices 
$\Sigma_{\pm}$ are equal, that is $\mathsf{K}_+=\mathsf{K}_-$ for 
corresponding quadrilaterals, then
the parallel lattices $\Sigma_1$ and~$\Sigma_2$ defined by
\bela{E49}
  \up{R}_1 = \up{R}_+ - \up{R}_-,\quad \up{R}_2 = \up{R}_+ + \up{R}_-
\end{equation}
constitute discrete isothermic surfaces which are related by the 
discrete Christoffel transformation.}

\medskip
\noindent
Conversely, any surface $\Sigma_1$ of constant discrete mean curvature
$\mathsf{M}_1=1$ 
constitutes a discrete isothermic surface and is associated with two particular
parallel discrete surfaces, namely a second surface $\Sigma_2$ of constant 
discrete mean curvature $\mathsf{M}_2=-1$ 
and a `middle' surface $\Sigma_+$ of constant positive
discrete Gau{\ss}ian curvature $\mathsf{K}_+=1$. 
It is interesting to note that the surfaces
$\Sigma_1,\Sigma_2$ and $\Sigma_+$ are at constant `distance', that is 
\mbox{$|\up{R}_1-\up{R}_+|=|\up{R}_2-\up{R}_+|=1$}, $|\up{R}_2-\up{R}_1|=2$.

\subsubsection{Discrete `linear' Weingarten surfaces} 

Surfaces of constant Gau{\ss}ian
or mean curvature represent particular examples of 
{\em Weingarten surfaces} (Eisenhart 1960), that is surfaces in 
$\mathbb{R}^3$ which admit
a functional relation between the principal curvatures. `Linear' Weingarten 
surfaces are those corresponding to a functional relation of the form
\bela{E49a}
  \alpha\mathsf{K} + \beta\mathsf{M} = \gamma,
\end{equation}
where $\alpha,\beta$ and $\gamma$ are arbitrary constants. If $\Sigma$ 
constitutes a {\em discrete linear Weingarten surface}, that is a curvature
lattice subject to the constraint (\ref{E49a}) with respect to a spherical
representation $\Sigma_{\circ}$, then, on use of the expressions (\ref{E17})
and~(\ref{E22}) for the discrete Gau{\ss}ian and mean 
curvatures $\mathsf{K}$ and $\mathsf{M}$ respectively, the above relation may 
be brought into the form
\bela{E49b}
  \down{H}_{(2)}\cdot\down{K} + 
  \down{K}_{(1)}\cdot\down{H} = 0,\quad S = \left(\bear{cc}\gamma & \beta\\
  \beta & -\alpha\ear\right)
\end{equation}
with the identification
\bela{E49c}
  \down{H} = (H,H_{\circ}),\quad\down{K} = (K,K_{\circ}).
\end{equation}
Thus, discrete linear Weingarten surfaces constitute discrete O surfaces which 
are parallel to surfaces of discrete constant Gau{\ss}ian curvature since the 
matrix $S$ as given by
(\ref{E49b})$_2$ may be mapped by means of an appropriate similarity 
transformation to either (\ref{E29}) or~(\ref{E32}) provided that 
$\det S\neq 0$. At the level of the position matrix $\both{R}$, this
corresponds to a linear transformation.

\subsubsection{Discrete Guichard surfaces} 

The class of discrete O surfaces in
$\mathbb{R}^3$
corresponding to a three-dimensional dual space endowed with 
the indefinite metric
\bela{E50}
  S = \left(\bear{ccc}0&1&0\\ 1&0&0\\ 0&0&1\ear\right)
\end{equation}
evidently includes discrete isothermic surfaces. On the other hand, in 
Schief \& Konopelchenko (2000), it has been shown how classical {\em
Guichard surfaces} (Eisenhart 1962) may be retrieved in the 
differential-geometric setting
by confining $\Sigma_3$ to the unit sphere. Accordingly, in the present
context, if 
$\Sigma_3$ is taken to be a spherical representation of $\Sigma_1$ and 
$\Sigma_2$ then the latter constitute {\em discrete Guichard
surfaces}.

\subsubsection{Discrete Petot surfaces} 

Another canonical class of discrete O surfaces 
is obtained by identifying the three-dimensional Euclidean space with its dual.
Thus, if we set
\bela{E53}
  \down{H} = \up{X}\tra,\quad \down{K} = \up{Y}\tra,\quad p=q
\end{equation}
then the linear systems (\ref{E24}) coincide and the discrete orthogonality
condition (\ref{E9})$_2$ reduces to
\bela{E54}
  \up{X}\cdot\up{Y} + p = 0,
\end{equation}
leading to
\bela{E55}
   \up{X}_{(2)}\cdot\up{Y} = 0,\quad \up{Y}_{(1)}\cdot\up{X} = 0.
\end{equation}
Alternatively, the identification
\bela{E56}
  \down{H} = \up{X}\tra,\quad \down{K} = -\up{Y}\tra,\quad p=-q
\end{equation}
results in
\bela{E57}
  \up{X}\cdot\up{Y} = 0,\quad \up{X}_{(2)}\cdot\up{Y}_{(1)} = 0.
\end{equation}
The characterizations (\ref{E55}) or (\ref{E57}), which, in geometric terms,
express the fact that there exist two right angles in each quadrilateral,
have been used to define {\em discrete Petot surfaces} (Schief 1997).
These represent the constituent members of discrete Darboux-Egorov-type
triply orthogonal systems of surfaces. It is noted, however, that the above
discrete Petot O surfaces are not generic due to the particular 
form~(\ref{E53})$_{1,2}$ or~(\ref{E56})$_{1,2}$ of 
the adjoint eigenfunctions $\down{H}$ and~$\down{K}$.
The class of discrete Petot O surfaces nevertheless enshrines the generic
class of discrete Petot surfaces in the sense that any discrete Petot
surface may be obtained from a discrete Petot O~surface by application of an
appropriate discrete Combescure transformation. 

\section{A B\"acklund transformation for discrete O surfaces}

The transformation theory of conjugate lattices is by now well-established
(Kono\-pel\-chenko \& Schief 1998; Doliwa {\sl et al.\ }1997). 
Here, we focus on the
natural discrete analogue of the classical Fundamental transformation
(Eisenhart 1962). Since the {\em discrete Fundamental transformation}
(Konopelchenko \& Schief 1998) commutes with the discrete Combescure 
transformation, it may be simultaneously applied to sets of parallel
conjugate lattices.

\subsection{The discrete Fundamental and Ribaucour transformations}

The discrete Fundamental transformation
is generated by two pairs of scalar solutions of the linear 
systems~(\ref{E24}) and 
corresponding bilinear potentials of the form~(\ref{E25}). Thus, for a 
given pair of functions $p,q$ associated with a set of parallel conjugate 
lattices $\Sigma_{\kappa}$, let $\{X,Y\}$ and $\{H,K\}$ be solutions
of the linear systems
\bela{E58}
  \bear{rlrl}
    X_{(2)} = & \dis\frac{X+qY}{\Gamma},\quad & 
    H_{(2)} = & \dis\frac{H+pK}{\Gamma}\AS
    Y_{(1)} = & \dis\frac{Y+pX}{\Gamma},\quad & 
    K_{(1)} = & \dis\frac{K+qH}{\Gamma}.
  \ear
\end{equation}
In the sequel, we refer to $X,Y$ and $H,K$ as {\em eigenfunctions} and {\em 
adjoint eigenfunctions} respectively. Three bilinear potentials 
$\up{M},\down{M}$ and $M$ may now be introduced according to
\bela{E59}
 \bear{rlrlrl}
   \up{M}_{(1)}-\up{M} = & \up{X}H,\quad & 
   \down{M}_{(1)}-\down{M} = & X\down{H},\quad & 
   M_{(1)}-M = & XH\as
   \up{M}_{(2)}-\up{M} = & \up{Y}K,\quad & 
   \down{M}_{(2)}-\down{M} = & Y\down{K},\quad & 
   M_{(2)}-M = & YK.
 \ear
\end{equation}
A second set of parallel conjugate lattices $\Sigma_{\kappa}'$ is now 
obtained as follows (Konopel\-chenko \& Schief 1998):

\begin{theorem}\label{fundamental}
{\bf (The discrete Fundamental transformation)} 
The linear systems (\ref{E24}) and the
defining relations (\ref{E25}) are invariant under
\bela{E60}
  (\both{R},\up{X},\up{Y},\down{H},\down{K},p,q)\rightarrow
  (\both{R}',\up{X}',\up{Y}',\down{H}',\down{K}',p',q'),
\end{equation}
where
\bela{E61}
  \both{R}' = \both{R} - \frac{\up{M}\down{M}}{M}
\end{equation}
and
\bela{E62}
 \bear{rlrl}
   \up{X}' = & \dis \sqrt{\frac{M}{M_{(1)}}}
                    \left(\up{X} - \frac{X\up{M}}{M}\right),\quad & 
   \up{Y}' = & \dis \sqrt{\frac{M}{M_{(2)}}}
                    \left(\up{Y} - \frac{Y\up{M}}{M}\right)\AS
   \down{H}' = & \dis \sqrt{\frac{M}{M_{(1)}}}
                      \left(\down{H} - \frac{H\down{M}}{M}\right),\quad &
   \down{K}' = & \dis \sqrt{\frac{M}{M_{(2)}}}
                      \left(\down{K} - \frac{K\down{M}}{M}\right)\AS
   p' = & \dis\sqrt{\frac{M^2}{M_{(1)}M_{(2)}}}
              \left(p - \dis \frac{YH}{M}\right),\quad & 
   q' = & \dis\sqrt{\frac{M^2}{M_{(1)}M_{(2)}}}
              \left(q - \dis \frac{XK}{M}\right).
 \ear
\end{equation}
\end{theorem}
\noindent  
In order to verify the above transformation laws, it is convenient to be
aware of the relation
\bela{E62a}
  \Gamma^{\prime2} = \frac{MM_{(12)}}{M_{(1)}M_{(2)}}\,\Gamma^2.
\end{equation}
It is emphasized that the above transformation may also be regarded as a 
mapping between the sets of discrete dual surfaces $\Sigma^k$ and 
$\Sigma^{\prime k}$. 

If the conjugate lattices $\Sigma_{\kappa}$ are discrete orthogonal, that
is the condition (\ref{E9})$_2$ is satisfied, then it is readily verified that
the quantities
\bela{E63}
  X = (\up{M}_{(1)}+\up{M})\cdot\up{X},\quad 
  Y = (\up{M}_{(2)}+\up{M})\cdot\up{Y}
\end{equation}
constitute particular eigenfunctions. This choice of eigenfunctions in the
definitions of the bilinear potentials $\up{M}$ and $M$ leads, in turn, to
the relations
\bela{E64}
  \up{M}_{(1)}^2-\up{M}^2 = M_{(1)}-M,\quad 
  \up{M}_{(2)}^2-\up{M}^2 = M_{(2)}-M
\end{equation}
so that we may set
\bela{E65}
  \up{M}^2 = M.
\end{equation}
It is now straightforward to show that
\bela{E66}
  \up{X}^{\prime2} = 1,\quad 2\up{X}'\cdot\up{Y}' + p' + q' = 0,\quad
  \up{Y}^{\prime2} = 1.
\end{equation}
Thus, it turns out that curvature lattices and the normalization
(\ref{E11}) are preserved by the discrete Fundamental transformation if the
eigenfunctions $X,Y$ and the bilinear potential~$M$ are chosen to be
(\ref{E63}) and (\ref{E65}) respectively. Under these circumstances, the
discrete Fundamental transformation becomes the {\em discrete \mbox{Ribaucour} 
transformation} as set down in Konopelchenko \& Schief (1998). Indeed, in the 
natural continuum limit, the above particular discrete Fundamental 
transformation reduces to the classical Ribaucour transformation for 
surfaces in $\mathbb{R}^3$ parametrized in terms of curvature 
coordinates~(Eisenhart 1962).

\subsection{Application to discrete O surfaces}

It is remarkable that the discrete Ribaucour transformation may be constrained
in such a way that discrete orthogonality of the dual conjugate lattices  
is also sustained. In fact, as a by-product, a parameter-dependent linear 
representation of discrete O~surfaces is obtained. As in the preceding, we 
first observe that the quantities
\bela{E67}
  H = \lambda(\down{M}_{(1)}+\down{M})\cdot\down{H},\quad 
  K = \lambda(\down{M}_{(2)}+\down{M})\cdot\down{K}
\end{equation}
constitute particular adjoint eigenfunctions. The constant parameter 
$\lambda$ is now non-trivial as we have already specified the eigenfunctions 
$X$ and $Y$. The associated potentials $\down{M}$ and $M$ then obey the 
relations
\bela{E68}
  \lambda(\down{M}_{(1)}^2-\down{M}^2) = M_{(1)}-M,\quad
  \lambda(\down{M}_{(2)}^2-\down{M}^2) = M_{(2)}-M
\end{equation}
so that
\bela{E69}
  \lambda\down{M}^2 = M
\end{equation}
is, at least, consistent. It is shown below that this constraint is indeed
admissible. On this assumption, we now proceed and note that
\bela{E70}
  \down{H}^{\prime2} = \down{H}^2,\quad 
 2\down{H}'\cdot\down{K}' + q'\down{H}^{\prime2} 
 + p'\down{K}^{\prime2} = 0,\quad
  \down{K}^{\prime2} = \down{K}^2,
\end{equation}
which implies that the transformed dual conjugate lattices $\Sigma^{\prime k}$
are also discrete orthogonal with any normalization of the form
\bela{E70a}
  \down{H}^2 = \alpha(n_1),\quad \down{K}^2 = \beta(n_2)
\end{equation}
unchanged (cf.\ (\ref{E28})).
 
Finally, insertion of the (adjoint) eigenfunctions $X,Y$ and $H,K$ as given
by (\ref{E63}) and (\ref{E67}) respectively into the defining relations
(\ref{E59}) produces the following {\em Lax pair} for discrete O surfaces:

\begin{theorem}\label{lax}
{\bf (A Lax pair for discrete O surfaces)} The linear system
\bela{E71}
 \bear{rl}
  \left(\bear{c}\up{M}\\ \down{M}\tra\ear\right)_{(1)}
- \left(\bear{c}\up{M}\\ \down{M}\tra\ear\right) = &
  \dis\frac{2}{1-\alpha\lambda}
  \left(\bear{cc}\alpha\lambda\up{X}\up{X}\tra&\lambda\up{X}\down{H}\\ 
                 \down{H}\tra\up{X}\tra&\lambda\down{H}\tra\down{H}\ear\right)
  \left(\bear{c}\up{M}\\ \down{M}\tra\ear\right)
  \\[5mm]
  \left(\bear{c}\up{M}\\ \down{M}\tra\ear\right)_{(2)} 
- \left(\bear{c}\up{M}\\ \down{M}\tra\ear\right) = &
  \dis\frac{2}{1-\beta\lambda}
  \left(\bear{cc}\beta\lambda\up{Y}\up{Y}\tra&\lambda\up{Y}\down{K}\\ 
                 \down{K}\tra\up{Y}\tra&\lambda\down{K}\tra\down{K}\ear\right)
  \left(\bear{c}\up{M}\\ \down{M}\tra\ear\right)
 \ear
\end{equation}
is compatible modulo the linear systems (\ref{E24}) and the discrete 
orthogonality conditions (\ref{E9}) and (\ref{E27}). It admits the first 
integral
\bela{E72}
  \up{M}^2 - \lambda\down{M}^2 = \mbox{\rm const}.
\end{equation}
\end{theorem}

\noindent
The existence of the first integral (\ref{E72}) guarantees that the constraint 
(\ref{E69}) is admissible. Consequently, we are now in a position to formulate 
the following theorem:

\begin{theorem}\label{back}
{\bf (A B\"acklund transformation for discrete O surfaces)} Let $\both{R}$ be 
the position matrix of a set of parallel discrete O surfaces 
$\Sigma_{\kappa}$ and their duals $\Sigma^k$ and 
$\up{X},\up{Y},\down{H},\down{K}$ corresponding
tangent vectors. If the vectors $\up{M}$ and $\down{M}$ constitute a solution
of the linear system (\ref{E71}) subject to the admissible constraint
\bela{E73}
  \up{M}^2 = \lambda\down{M}^2 = M
\end{equation}
and the scalar $M$ is defined by the latter then the position matrix of a 
second set of discrete O surfaces $\Sigma'_{\kappa},\Sigma^{\prime k}$ is 
given by
\bela{E74}
  \both{R}' = \both{R} - \frac{\up{M}\down{M}}{M}.
\end{equation}
\end{theorem}

We conclude this section with the following important observation.
If a discrete O~surface $\Sigma_n$
is identified with a spherical representation of the remaining discrete 
O~surfaces $\Sigma_{\kappa}$ then
\bela{E75}
  \up{R}_n^2 = 1 
\end{equation}
and it is readily verified that
\bela{E76}
  M_n = 2\up{R}_n\cdot\up{M}
\end{equation}
is another admissible constraint. Consequently, the $n$th component of the
transformation law (\ref{E74}) may be cast into the form
\bela{E77}
  \up{R}'_n = \left(\eins - 2\frac{\up{M}\up{M}\tra}{\up{M}^2}\right)\up{R}_n
\end{equation}
which implies that
\bela{E78}
  \up{R}^{\prime2}_n = 1.
\end{equation}
Hence, we come to the important conclusion that the above B\"acklund 
transformation acts within specific sub-classes of discrete O surfaces such
as surfaces of constant discrete Gau{\ss}ian curvature, discrete minimal or 
discrete Guichard surfaces. Moreover, it is readily shown that constraints of
the form
\bela{E78a}
  \left(\sum_{\kappa=1}^n c_{\kappa}\up{R}_{\kappa}\right)^2=1,
\end{equation}
which generalize (\ref{E75}), may also be preserved. In particular, the
specialization (\ref{E46}) leading to constant discrete mean curvature 
surfaces proves invariant.

\section{The discrete pseudosphere and discrete breather pseudospherical
surfaces}

We conclude this paper with an illustration of the B\"acklund transformation
for discrete O~surfaces and consider the particular case (\ref{E29}) of 
discrete pseudospherical surfaces. Thus, we here regard a straight
polygon as a (degenerate) seed discrete pseudospherical surface $\Sigma_1$ 
together with an associated `spherical representation'~$\Sigma_2$ 
represented~by
\bela{E79}
  \up{R}_{1} = \left(\bear{c}0\\ 0\\ \epsilon n\ear\right),\quad
  \up{R}_{2} = \left(\bear{c}-\sin \nu m\\ 
                                       \cos \nu m\\ 0\ear\right)
\end{equation}
so that the tangent vectors to $\Sigma_1,\Sigma_2$ and their duals read
\bela{E80}
  \up{X} = \left(\bear{c}0\\ 0\\ 1\ear\right),\quad
  \up{Y} = \left(\bear{c}\cos(\nu m + \nu/2)\\ 
                         \sin(\nu m + \nu/2)\\ 0\ear\right),\quad
  \down{H} = (\epsilon\,\,\,0),\quad \down{K} = (0\,\,\,-\delta)
\end{equation}
with
\bela{E80a}
  (n,m) = (n_1,n_2),\quad \delta = 2\sin\nu/2.
\end{equation}
It is evident that the linear systems (\ref{E24}) with $p=q=0$ and the 
orthogonality conditions $\up{X}\cdot\up{Y}=\down{H}\cdot\down{K}=0$ are
satisfied. Accordingly, the linear system (\ref{E71}) for these particular
discrete O surfaces becomes
\bela{E81}
 \bear{rl}
  \left(\bear{c}M^1\\ M^2\\ M^3\\ M_1\\ M_2\ear\right)_{(1)}-
  \left(\bear{c}M^1\\ M^2\\ M^3\\ M_1\\ M_2\ear\right) = &\dis
  \frac{2\epsilon}{1-\epsilon^2\lambda}
  \left(\bear{ccccc}0&0&0&0&0\\ 0&0&0&0&0\\ 0&0&\lambda\epsilon&\lambda&0\\
                    0&0&1&\lambda\epsilon&0\\ 0&0&0&0&0\ear\right)
  \left(\bear{c}M^1\\ M^2\\ M^3\\ M_1\\ M_2\ear\right)\\[12mm]
  \left(\bear{c}M^1\\ M^2\\ M^3\\ M_1\\ M_2\ear\right)_{(2)} - 
  \left(\bear{c}M^1\\ M^2\\ M^3\\ M_1\\ M_2\ear\right) = &\dis
  \frac{2\delta}{1-\delta^2\lambda}
  \left(\bear{ccccc}
      \lambda\delta (Y^1)^2&\lambda\delta Y^1Y^2&0&0&-\lambda Y^1\\ 
      \lambda\delta Y^1Y^2&\lambda\delta(Y^2)^2&0&0&-\lambda Y^2\\ 
                    0&0&0&0&0\\ 0&0&0&0&0\\ 
                    -Y^1&-Y^2&0&0&\lambda\delta\ear\right)
  \left(\bear{c}M^1\\ M^2\\ M^3\\ M_1\\ M_2\ear\right).
 \ear
\end{equation}
The latter decouples into two systems of linear {\em non-autonomous} ordinary 
difference
equations for $M^3(n),M_1(n)$ and $M^1(m),M^2(m),M_2(m)$ respectively.
The constants of integration in the general solution of (\ref{E81}) have to 
be chosen in such a way that the admissible constraints (\ref{E73}) and 
(\ref{E76})$_{n=2}$ are satisfied. 

\begin{figure}[h]
 \begin{minipage}[t]{0.49\textwidth}
 \centerline{\includegraphics[angle=270,trim= 110 190 110 190,width=\textwidth,
                                               clip]{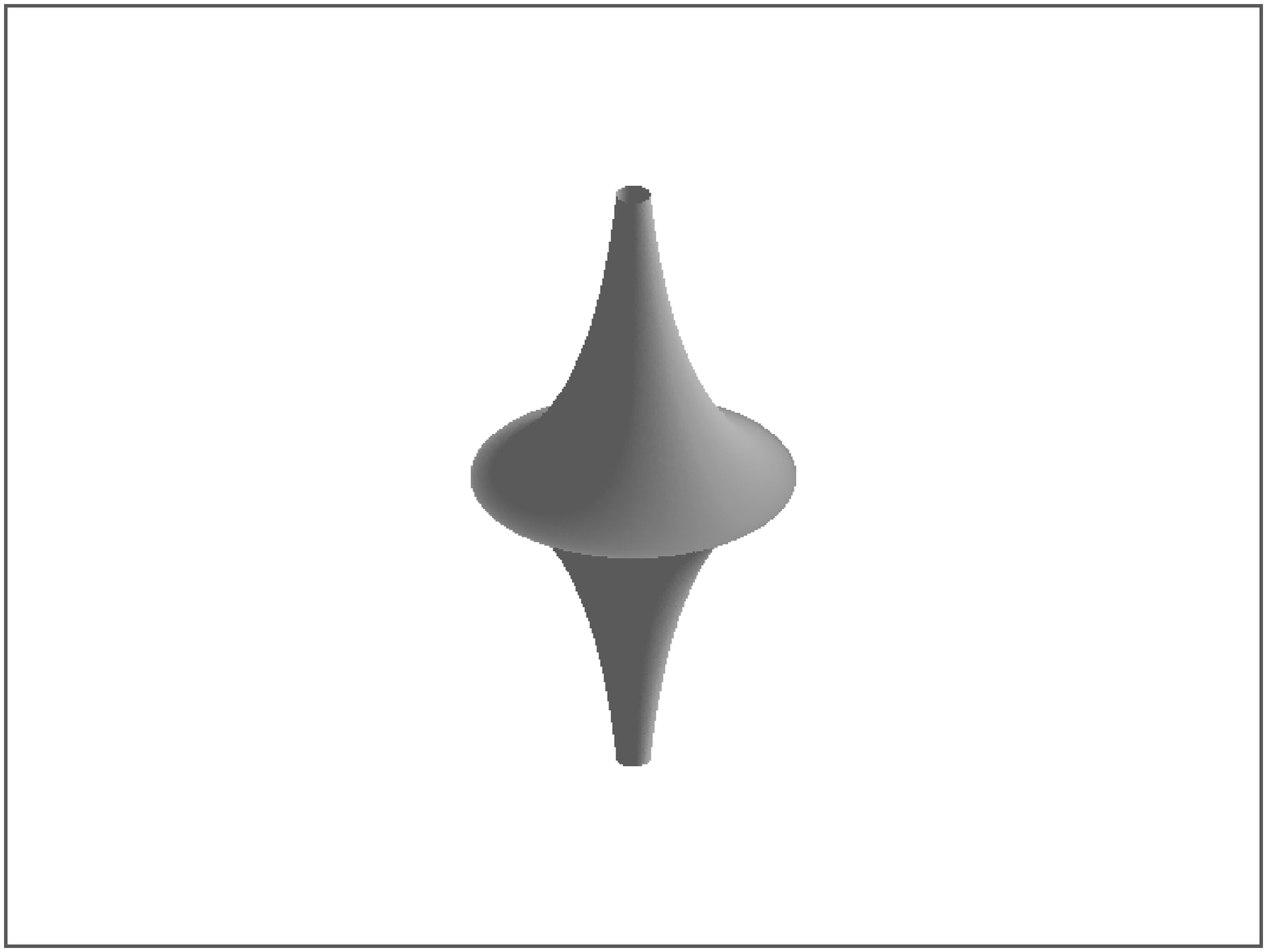}}
 \caption{The classical pseudosphere}
 \label{dummy}
 \end{minipage}
 \begin{minipage}[t]{0.49\textwidth}
 \centerline{\includegraphics[angle=270,trim= 110 190 110 190,width=\textwidth,
                                   clip]{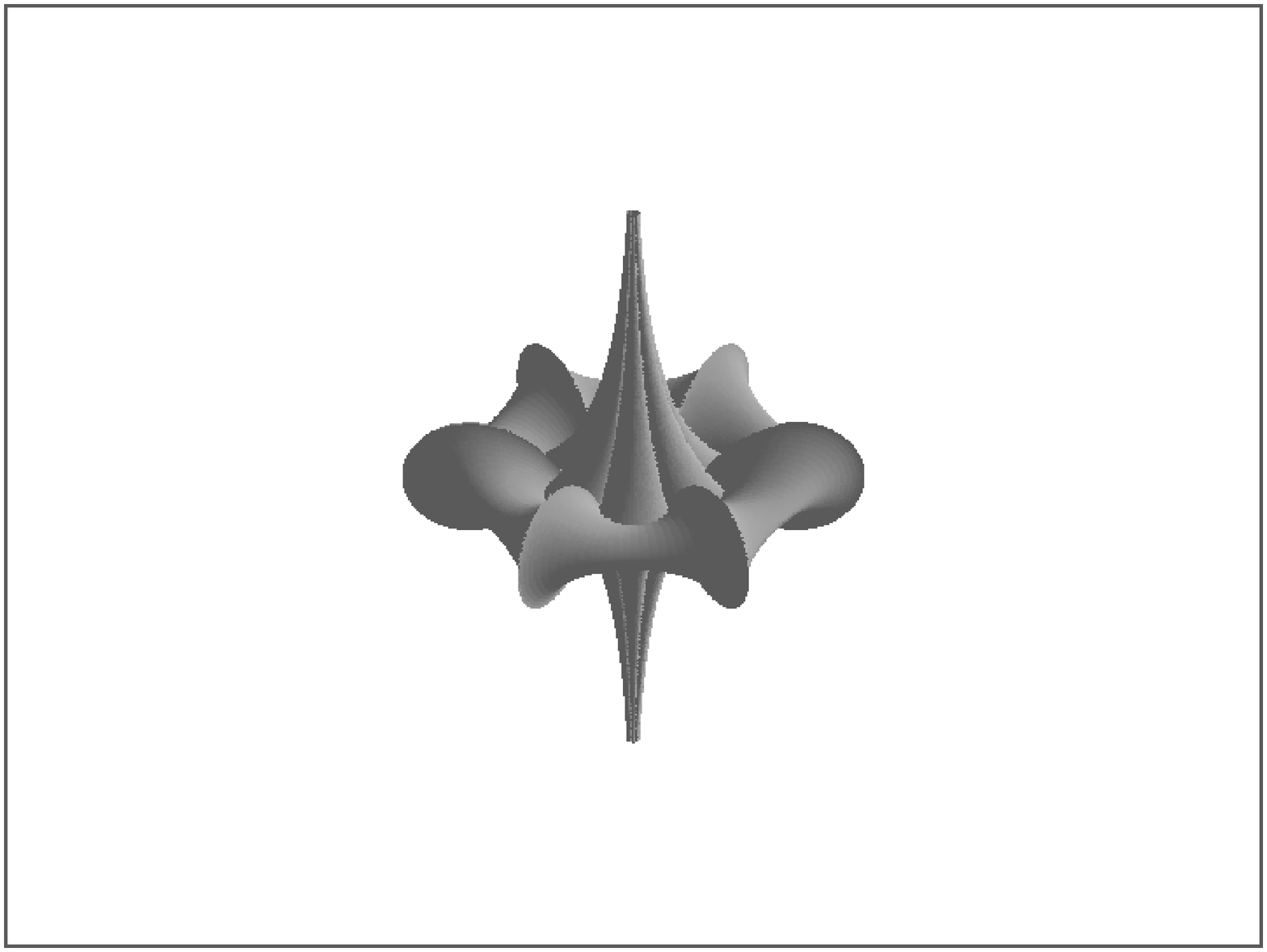}}
 \caption{A breather pseudospherical surface}
 \label{pseudo}
 \end{minipage}
\end{figure}

\begin{figure}[h]
 \begin{minipage}[t]{0.49\textwidth}
 \centerline{\includegraphics[angle=270,trim= 110 200 110 200,width=\textwidth,
                                               clip]{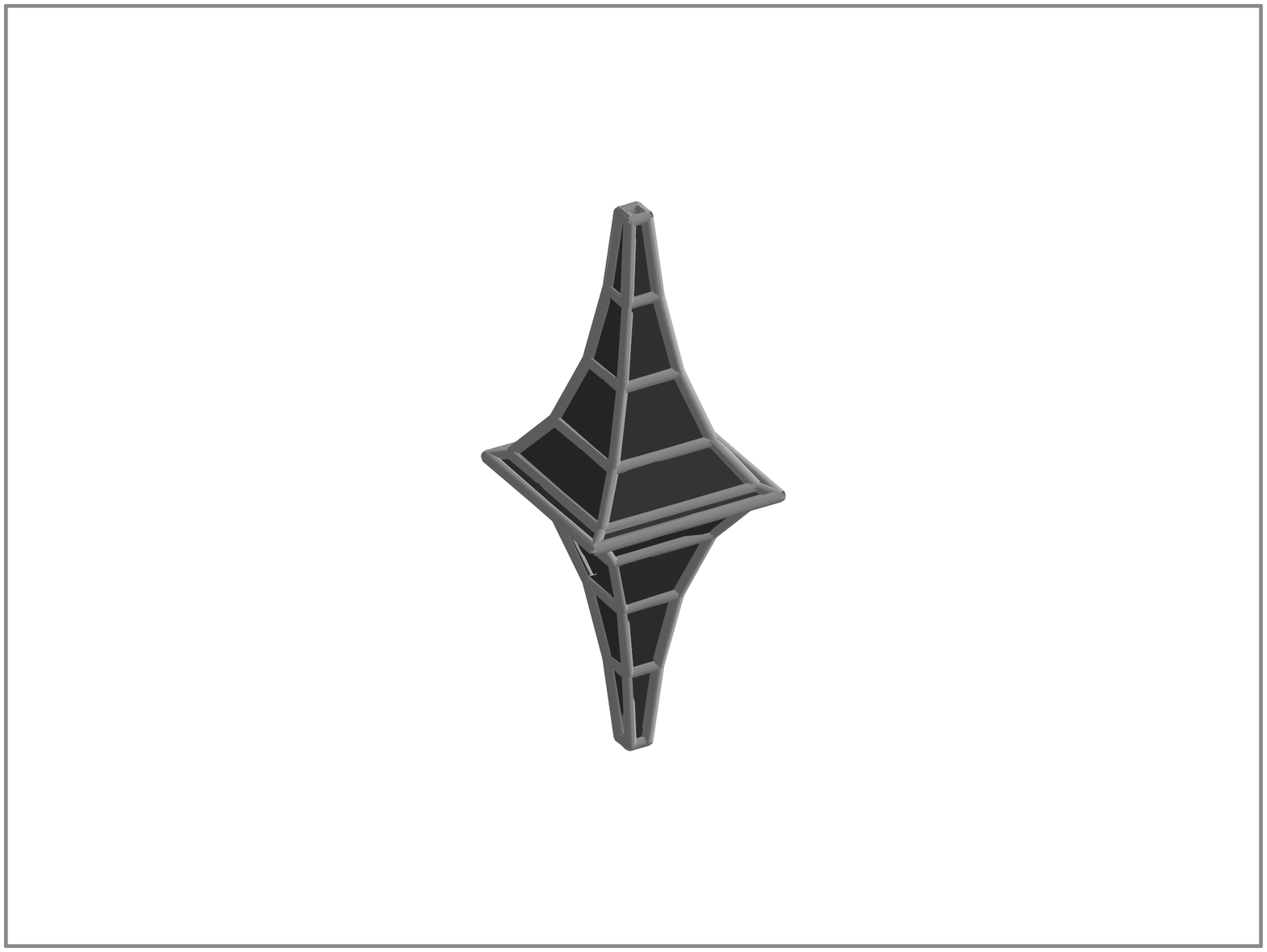}}
 \end{minipage}
 \begin{minipage}[t]{0.49\textwidth}
 \centerline{\includegraphics[angle=270,trim= 90 180 110 180,width=\textwidth,
                                   clip]{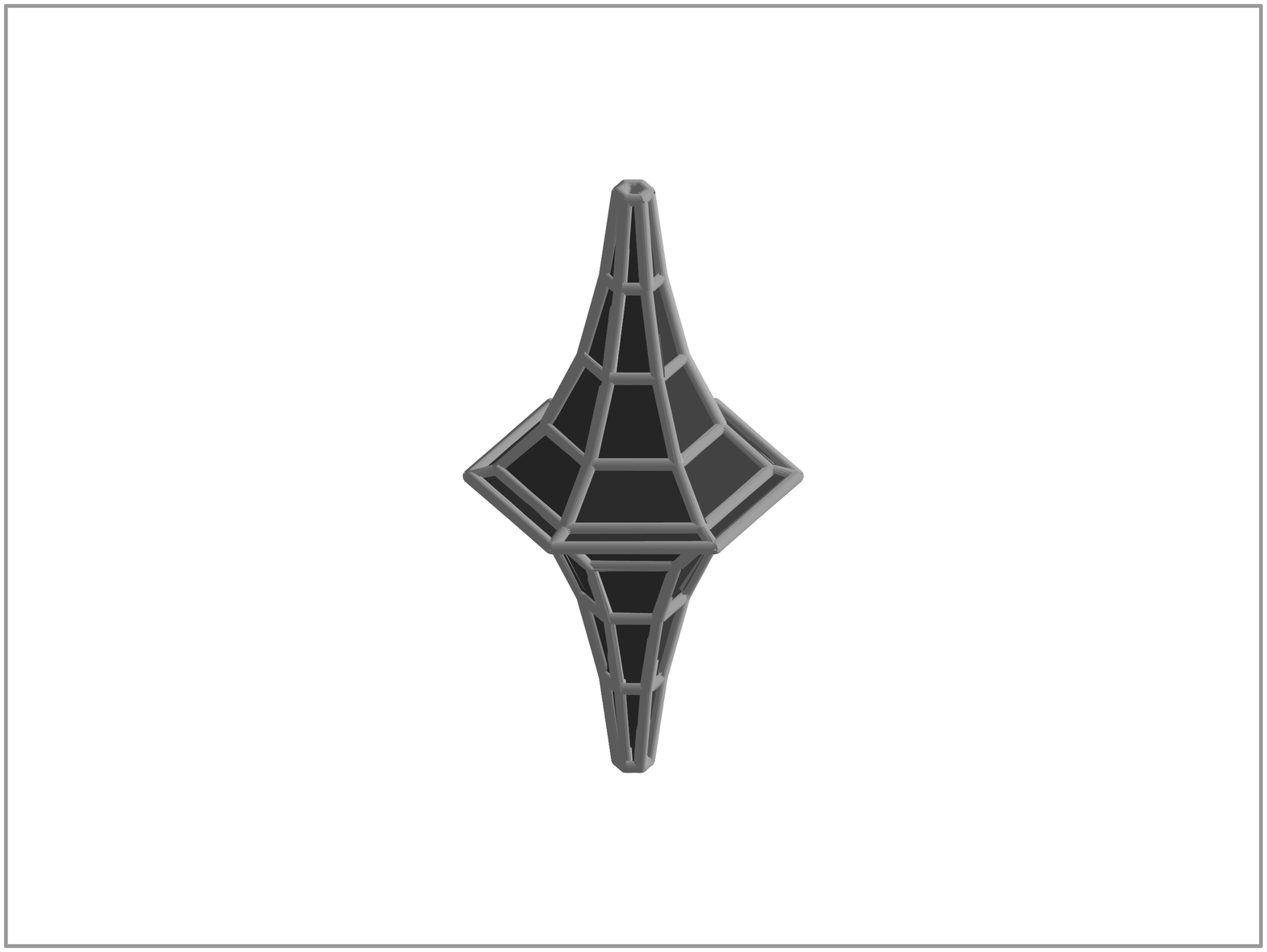}}
 \end{minipage}
 \caption{Discrete pseudospheres}
 \label{spheredisc}
\end{figure}

In the differential-geometric context, that is in the natural continuum limit
\bela{E81a}
  x = \epsilon n,\quad y = \nu m,\quad \epsilon,\nu\rightarrow0,
\end{equation}
it has been shown (Schief \& Konopelchenko 2000) that the 
B\"acklund transformation for O surfaces produces either Beltrami's classical
{\em pseudosphere} (Eisenhart 1960) or `stationary' 
{\em breather pseudopsherical
surfaces} (Rogers \& Schief 2000). The pseudosphere and a particular
breather pseudospherical surface with $\mathbb{Z}_6$ rotational symmetry
is depicted in Figure \ref{pseudo}.
It is therefore expected that the B\"acklund transformation for discrete
O surfaces delivers discretizations of the pseudosphere and breather
pseudospherical surfaces. For brevity, we here only state that a careful
analysis of the solution of the linear system (\ref{E81}) and
the associated B\"acklund transformation leads to the following result:

In the case $\lambda=1/4$, the position vector $\up{R}'_1$ of the B\"acklund 
transform $\Sigma'_1$ may be reduced to
\bela{E82}
  \up{R}'_1 = \left(\bear{c}\dis\frac{\sin \nu m}{\cosh \tau n}\AS
                         -\dis\frac{\cos \nu m}{\cosh \tau n}\As
                         \epsilon n - \tanh \tau n\ear\right)
\end{equation}
modulo translations of the form $n\rightarrow n+\mbox{const},\,m\rightarrow
m+\mbox{const}$, where
\bela{E82a}
  \tau = \ln\left(\frac{2+\epsilon}{2-\epsilon}\right).
\end{equation}
This discrete pseudsospherical surface of `revolution' is nothing but
a discretization of Beltrami's classical pseudosphere. The parameters
$\epsilon$ and $\nu$, which may be chosen arbitrarily, constitute measures of 
the `quality' of the discretization. If $\nu$ is rational then the discrete
pseudosphere admits a discrete rotational symmetry. Two such discrete
pseudospheres are displayed in Figure \ref{spheredisc}. 

\begin{figure}[h]
 \begin{minipage}{0.49\textwidth}
 \centerline{\includegraphics[angle=270,trim= 110 190 110 190,width=\textwidth,
                              clip]{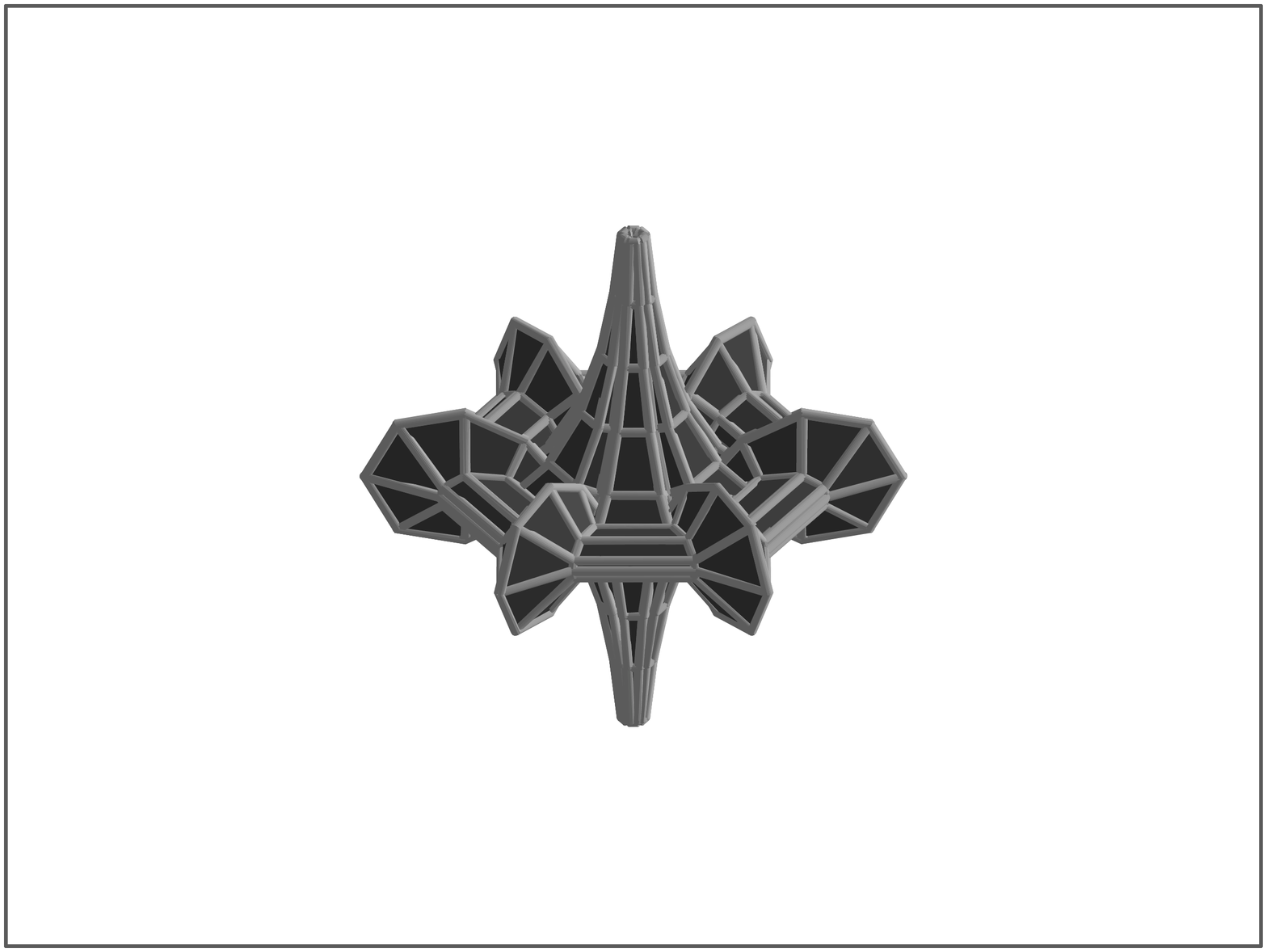}}
 \end{minipage}
 \begin{minipage}{0.49\textwidth}
 \centerline{\includegraphics[angle=270,trim= 110 190 110 190,width=\textwidth,
                              clip]{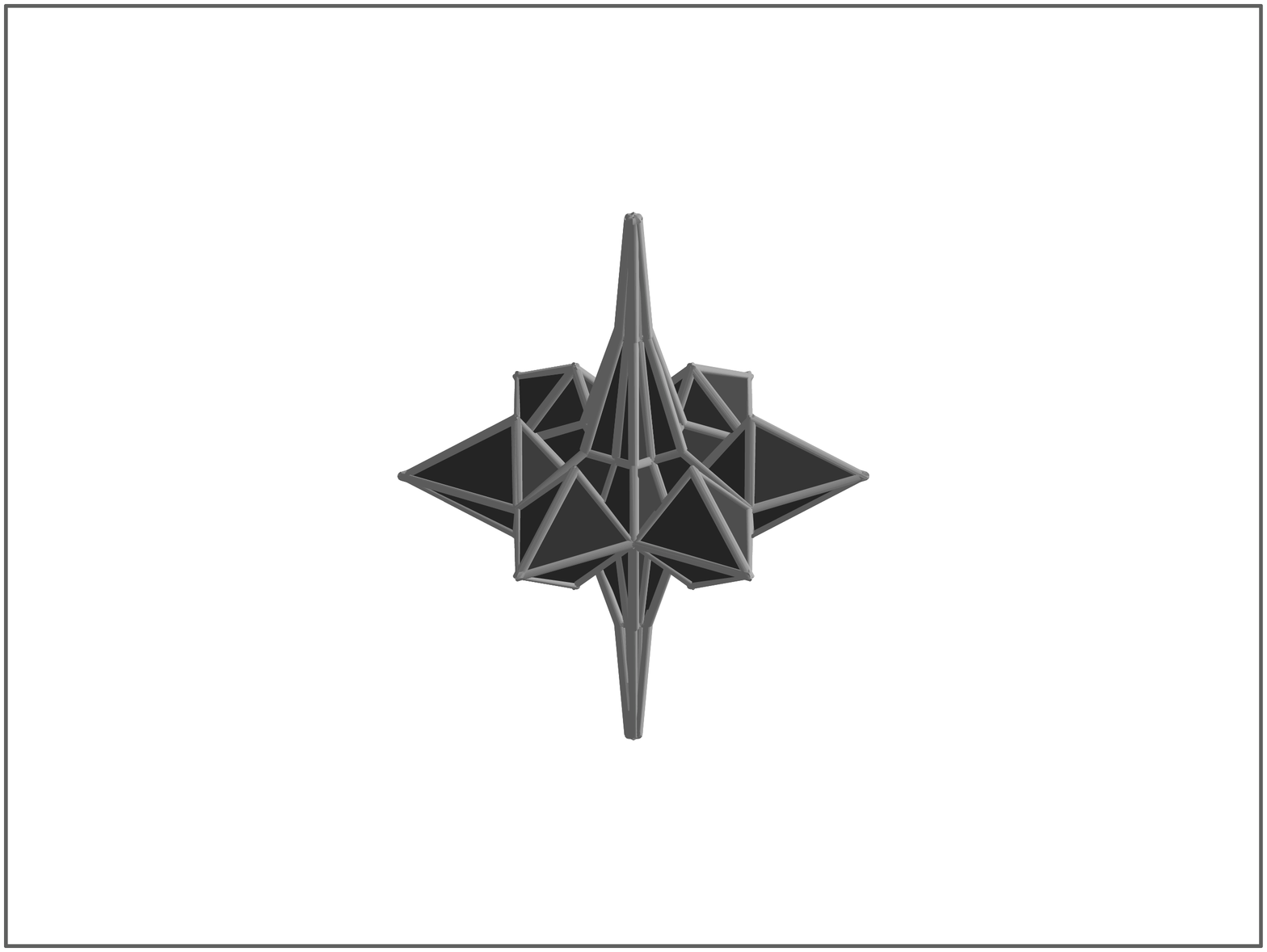}}
 \end{minipage}\\[2mm]
 \begin{minipage}{0.49\textwidth}
 \centerline{\includegraphics[angle=270,trim= 110 190 110 190,width=\textwidth,
                              clip]{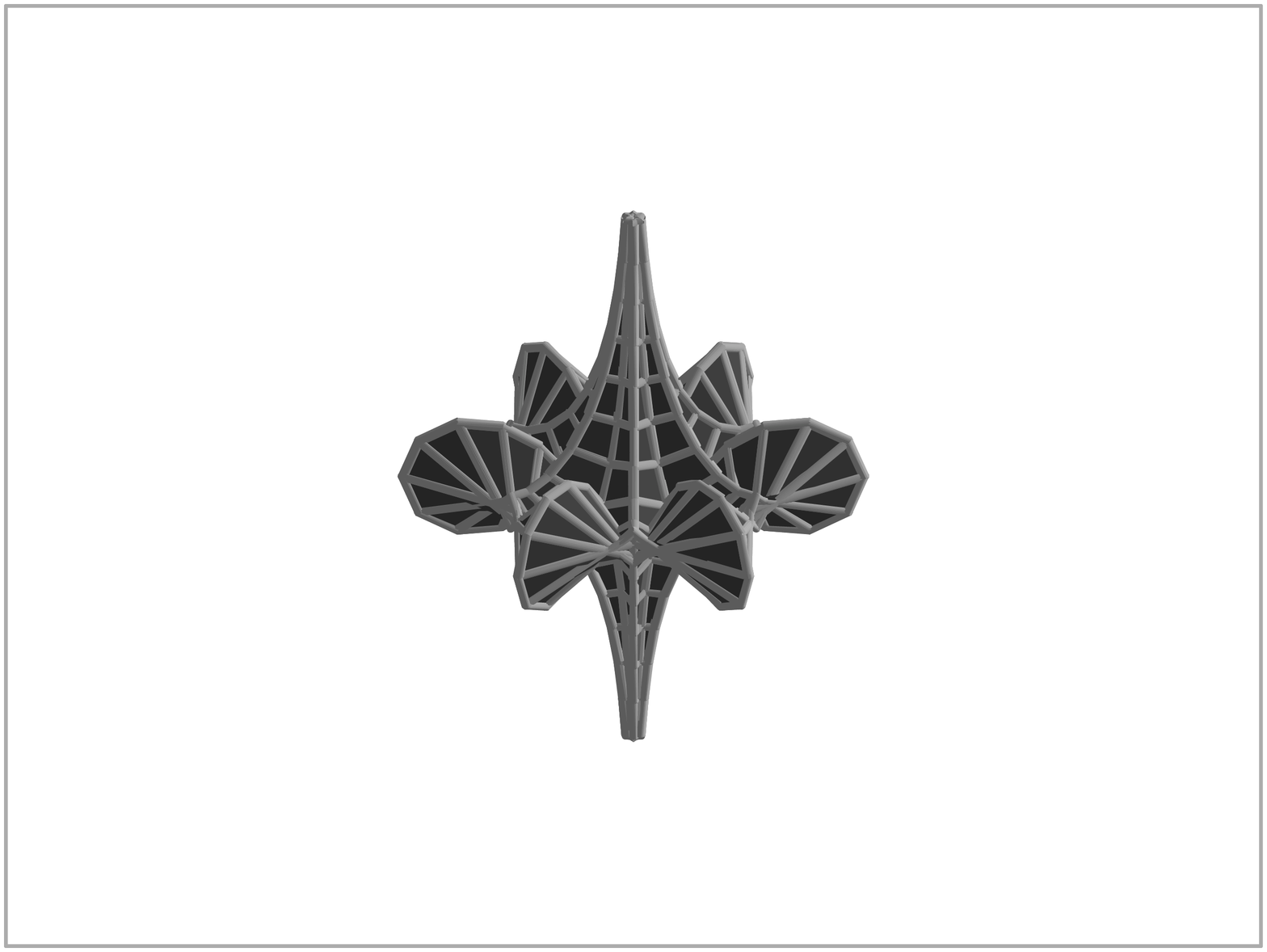}}
 \end{minipage}
 \begin{minipage}{0.49\textwidth}
 \centerline{\includegraphics[angle=270,trim= 110 190 110 190,width=\textwidth,
                              clip]{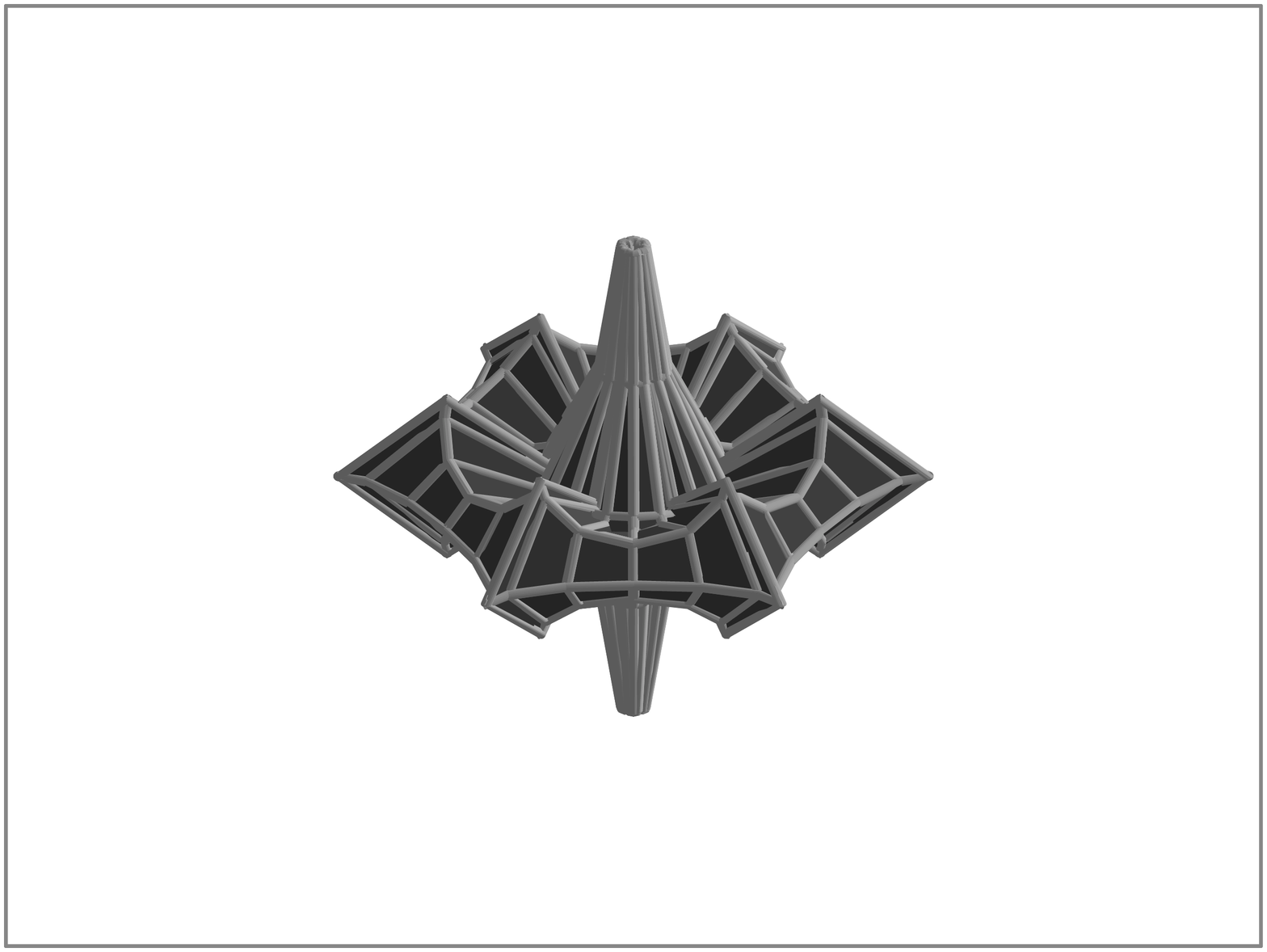}}
 \end{minipage}
 \caption{Discrete breather pseudospherical surfaces for  
     $\frac{\mathsf{p}}{\mathsf{q}} = \frac{3}{4}$}
 \label{breatherdisc}
\end{figure}

If $\lambda\neq 1/4$ then integration of the linear system (\ref{E81}) and 
specification of the constants of integration lead to the
position vector
\bela{E83}
 \bear{rl}
  \up{R}'_1 = \left(\bear{c}0\\ 0\\ \epsilon n\ear\right) - &\dis\frac{2d}{c}
     \frac{\cosh\tau n}{c^2\sin^2\kappa m + d^2\cosh^2\tau n}\times\\[6mm]
             &\left(\bear{c}-\sin\kappa m\sin\nu m - d\cos\kappa m\cos\nu m\\
                             \sin\kappa m\cos\nu m - d\cos\kappa m\sin\nu m\\ 
                             d\sinh\tau n\ear\right),
 \ear
\end{equation}
where
\bela{84}
  \lambda = \frac{c^2}{4}, \quad c^2 + d^2 = 1,\quad 
  \tau = \ln\left(\frac{2+\epsilon c}{2-\epsilon c}\right),\quad
  \kappa = 2\arctan\left(d\tan\frac{\nu}{2}\right).
\end{equation}
These discrete pseudospherical surfaces indeed constitute discretizations of
the above-mentioned breather pseudospherical surfaces.
If the constants $c$ and $d$ are real then there exists a discrete rotational
symmetry if $\kappa/\nu$ is rational, that is
\bela{E84}
  \frac{\kappa}{\nu} = 
  \frac{\mathsf{p}}{\mathsf{q}},\quad \mathsf{p},\mathsf{q}\in\mathbb{Z}.
\end{equation}
A variety of {\em discrete breather pseudospherical surfaces} 
corresponding to different choices of $\mathsf{p}$ and $\mathsf{q}$ may now be
generated. In Figure \ref{breatherdisc}, several discretizations of the
breather pseudospherical surface displayed in Figure \ref{pseudo} characterized
by $\mathsf{p}/\mathsf{q}=3/4$ are shown.

As in the differential-geometric context, 
it is interesting to note that the B\"acklund transformation for discrete O 
surfaces does not reduce to a discrete version of the classical B\"acklund 
transformation for pseudospherical surfaces. In fact, as discussed above, a 
single application of the B\"acklund transformation for discrete O~surfaces
to a straight polygon produces discrete pseudospheres or
discrete breather pseudospherical surfaces. By 
contrast, a single application of the classical B\"acklund transformation
to a straight line 
results in a one-parameter family of {\em Dini surfaces} including the 
pseudosphere~(Eisenhart 1960). 
A second application then leads to breather pseudospherical
surfaces if one assumes that the two B\"acklund parameters are complex 
conjugates (Rogers \& Schief 2000). 
It is also emphasized that the procedure for
the generation of discrete breather pseudospherical surfaces outlined here 
involves the solution of a system of {\em non-autonomous} difference equations.

\end{document}